\documentclass[aps,prl,twocolumn,superscriptaddress,amsmath,amssymb]{revtex4-1}
\usepackage[normalem]{ulem}
\usepackage{graphicx}
\usepackage{subfigure}
\usepackage{epsfig}
\usepackage{xcolor}
\usepackage{dcolumn}
\usepackage{bm}
\usepackage{ulem}
\usepackage{color}
\usepackage{amsmath}
\usepackage{amsfonts}
\usepackage{amssymb}

\begin{document}

\title{Evidence of charge density wave with anisotropic gap in monolayer VTe$_2$ film }

\author{Yuan Wang}
\altaffiliation{These authors contribute equally to this work.}
\affiliation{State Key Laboratory of Low Dimensional Quantum Physics and Department of Physics, Tsinghua University, Beijing 100084, China}

\author{Junhai Ren}
\altaffiliation{These authors contribute equally to this work.}
\affiliation{State Key Laboratory of Low Dimensional Quantum Physics and Department of Physics, Tsinghua University, Beijing 100084, China}
\affiliation{WPI-Advanced Institute for Materials Research, Tohoku University, Sendai 980-8577, Japan}

\author{Jiaheng Li}
\affiliation{State Key Laboratory of Low Dimensional Quantum Physics and Department of Physics, Tsinghua University, Beijing 100084, China}

\author{Yujia Wang}
\affiliation{State Key Laboratory of Low Dimensional Quantum Physics and Department of Physics, Tsinghua University, Beijing 100084, China}

\author{Huining Peng}
\affiliation{State Key Laboratory of Low Dimensional Quantum Physics and Department of Physics, Tsinghua University, Beijing 100084, China}

\author{Pu Yu}
\affiliation{State Key Laboratory of Low Dimensional Quantum Physics and Department of Physics, Tsinghua University, Beijing 100084, China}
\affiliation{Collaborative Innovation Center of Quantum Matter, Beijing, P.R. China}

\author{Wenhui Duan}
\affiliation{State Key Laboratory of Low Dimensional Quantum Physics and Department of Physics, Tsinghua University, Beijing 100084, China}
\affiliation{Collaborative Innovation Center of Quantum Matter, Beijing, P.R. China}

\author{Shuyun Zhou}
\altaffiliation{Correspondence should be sent to syzhou@mail.tsinghua.edu.cn}
\affiliation{State Key Laboratory of Low Dimensional Quantum Physics and Department of Physics, Tsinghua University, Beijing 100084, China}
\affiliation{Collaborative Innovation Center of Quantum Matter, Beijing, P.R. China}

\date{\today}

\begin{abstract}
{\bf We report experimental evidence of charge density wave (CDW) transition in monolayer 1T-VTe$_2$ film.  4$\times$4 reconstruction peaks are observed by low energy electron diffraction below the transition temperature $\textit{T}_\textup{CDW}$ = 186 K. Angle-resolved photoemission spectroscopy measurements reveal arc-like pockets with anisotropic
CDW gaps up to 50 meV. The anisotropic CDW gap is attributed to the imperfect nesting of the CDW wave vector, and first-principles calculations reveal phonon softening at the same vector, suggesting the important roles of Fermi surface nesting and electron-phonon interaction in the CDW mechanism.}
\end{abstract}

\maketitle

In layered transition-metal dichalcogenides (TMDCs) \cite{StranoNano2012,ZhangH2013,KisRev2017}, the electronic structure strongly depends on the sample thickness and quantum confinement effect can result in novel electronic properties and correlated phenomena, e.g. charge density wave (CDW) \cite{CDW1, CDW2, CDWRMP1988}, which are distinctive from the bulk materials. Vanadium dichalcogenide films have attracted intensive attention due to their intriguing properties, e.g., 2D magnetism and CDW.
Although bulk VSe$_2$ crystal is paramagnetic \cite{HaasSSC, SienkoJSSC},
room temperature ferromagnetism has been recently reported in monolayer VSe$_2$ film \cite{BatzilNatNano17,Gao18}.
Moreover, atomically thin VSe$_2$ films also exhibit rich CDW orders. While bulk VSe$_2$ crystal shows 4$\times4\times$3 CDW transition at 110 K \cite{VSe2bulk}, various CDW orders with different periods and transition temperatures have been reported for monolayer VSe$_2$ film, e.g., 4$\times$4 CDW order below 140 K \cite{KingNanoLetter18,SatoNanoResearch19}, $\sqrt{7}\times\sqrt{3}$ CDW below 220 K \cite{PRL18}, $\sqrt{7}\times\sqrt{3}$ and 2$\times\sqrt{3}$ CDW orders with transition temperatures of 135 K and 350 K, respectively \cite{ChangNanoLetter18}.

Compared to VSe$_2$, another vanadium dichalcogenide - VTe$_2$, is much less explored. Bulk VTe$_2$ shows more complicated phases with a structural phase transition from the high temperature 1T phase to the low temperature 1T$^\prime$ phase at 482 K, and CDW order of 3$\times1\times$3 has been suggested in the  1T$^\prime$ phase \cite{bronsema1984crystal,Nakamura2018}. Few-layer VTe$_2$ films, however, show 1T structure similar to VSe$_2$. Ferromagnetism has been reported in nanoplates of 1T-VTe$_2$ \cite{DuanAdvMat18}, and two possible CDW transitions at 240 K and 135 K have been reported in films of approximately ten layers thick from resistivity measurements \cite{Ma19}. However, the corresponding periods of those CDW orders remain unknown. Moreover, so far there is no report on the growth of monolayer VTe$_2$ film or its electronic structure. Resolving the CDW order in monolayer VTe$_2$ film and revealing the underlying mechanism are important.

Here we report the electronic structure of monolayer VTe$_2$ film grown by molecular beam epitaxy (MBE). Evidence of 4$\times$4 CDW ordering is reported in the low energy electron diffraction (LEED) below CDW transition temperature $\textit{T}_\textup{CDW}$  = 186 K. Angle-resolved photoemission spectroscopy (ARPES) measurements further reveal anisotropic CDW gaps arising from the V 3$\textit{d}$ orbital, forming arc-like Fermi surface. The anisotropic gaps can be explained by the imperfect nesting of the hole pockets. First-principles calculations of the phonon spectrum reveal phonon softening at the corresponding CDW wave vector, suggesting the important roles of electron-phonon interaction and Fermi surface nesting in the CDW mechanism.

\begin{figure*}
\centering
\includegraphics[width=17.8 cm] {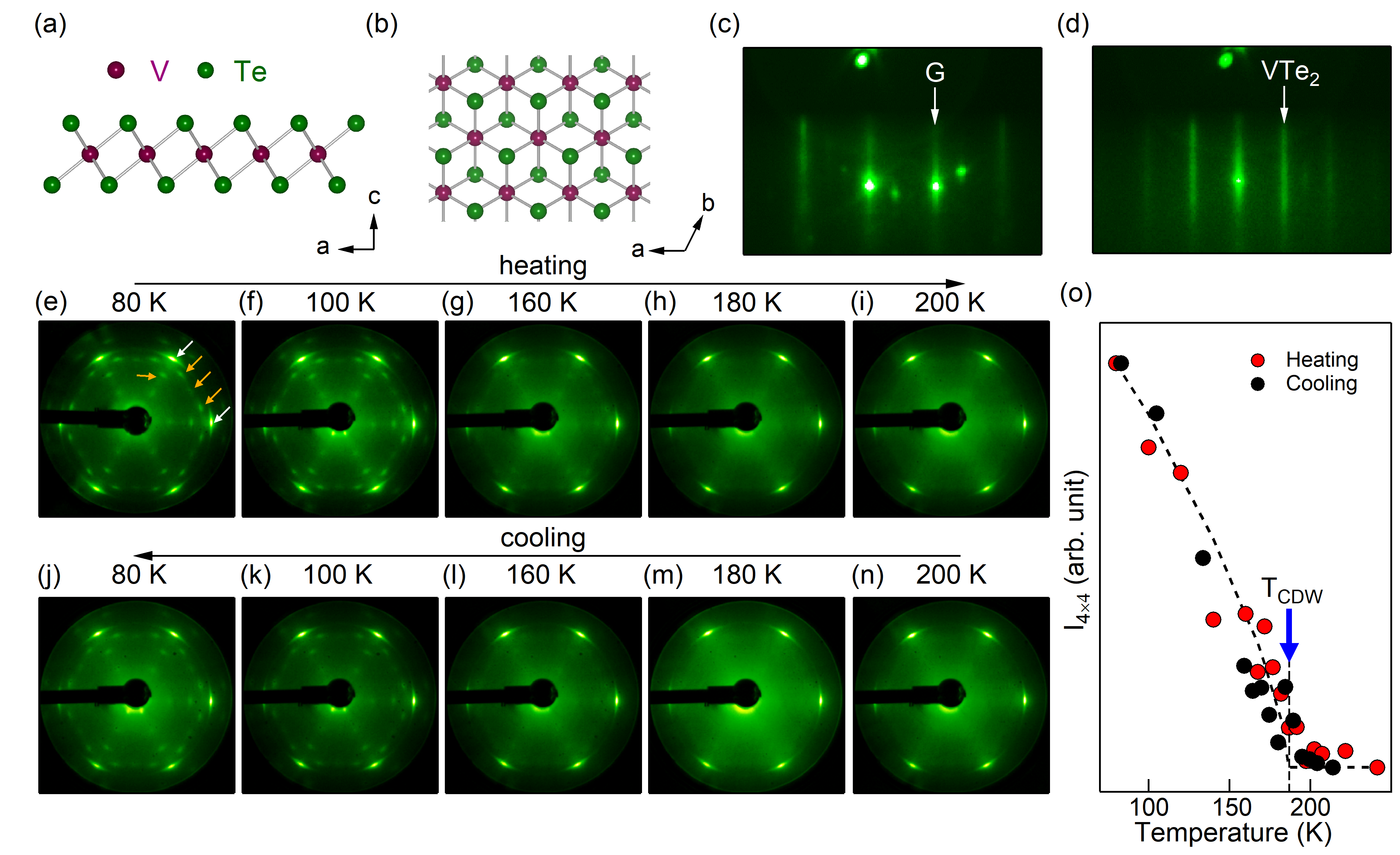}
\caption{ (a) and (b) Side (a) and top (b) views of the crystal structure of 1T-VTe$_2$. (c) and (d) RHEED patterns of BLG/SiC(0001) substrate (c) and the monolayer VTe$_2$ film (d). (e)-(i) Temperature dependent LEED patterns at a few selected temperatures in the warming process. The CDW peaks are indicated by yellow arrows in (e). (j)-(n) Temperature dependent LEED patterns in the cooling process. (o) Intensity of 4$\times$4 CDW peak normalized by VTe$_2$ Bragg peak during the warming (red symbols) and cooling (black symbols) processes, respectively. }
\label{Figure:1}
\end{figure*}

Monolayer 1T-VTe$_2$ consists of one layer of V atoms sandwiched between two layers of Te with space group $\textit{P}\bar{3}\textit{m}$1 (Figs.~\ref{Figure:1}(a) and (b) for the side and top views). Monolayer 1T-VTe$_2$ film is grown on bilayer graphene (BLG)/SiC(0001) substrate \cite{BLGsub} by MBE. The growth process is monitored by \textit{in situ} reflection high-energy electron diffraction (RHEED). Sharp patterns from the graphene substrate [Fig.~\ref{Figure:1}(c)] and 1T-VTe$_2$ film [Fig.~\ref{Figure:1}(d)] are observed in the RHEED patterns before and after growth of 1T-VTe$_2$ film. Using the lattice constant of graphene as a reference, the lattice constant of VTe$_2$ is extracted to be a = 3.63 \AA, which is similar to that of the nanoplate 3.64 \AA~\cite{DuanAdvMat18} and previous thicker MBE films 3.67 \AA~\cite{Ma19}. As shown in Fig.~S1 and Table~SI in the Supplemental Material \cite{Supplemental}, there is no significant change on lattice constant for 1T-VTe$_2$ film on BLG within a large growth temperature range, confirming the weak van der Waals interaction between the film and the substrate.

Figure \ref{Figure:1}(e) shows LEED pattern of monolayer VTe$_2$ film measured at 80 K. In addition to the Bragg spots from VTe$_2$ (pointed by white arrows), additional diffraction spots (pointed by yellow arrows) with 4$\times$4 period are observed, suggesting the CDW order of 4$\times$4. Figures 1(e)-1(i) show the evolution of the LEED patterns with increasing temperature. The intensity of 4$\times$4 diffraction spots decreases with temperature and fully disappears around 180 K, suggesting that the CDW transition occurs near this temperature. When cooling down below 180 K [Figs.~1(n)-(j)], the 4$\times$4 diffraction peaks are observed again, confirming that the 4$\times$4 reconstruction is an intrinsic temperature effect associated with the CDW transition. To quantify the CDW transition temperature, we show in Fig.~\ref{Figure:1}(o) the normalized intensity of 4$\times$4 peak as a function of temperature. By tracking the temperature-dependent LEED measurements, we determine the CDW transition temperature to be $\textit{T}_\textup{CDW}\approx$ 186 K. Similar CDW with weaker 4$\times$4 period has been reported in monolayer VSe$_2$ film below $\textit{T}_\textup{CDW}$=140 K \cite{KingNanoLetter18}.

\begin{figure*}
\centering
\includegraphics[width=17.8 cm] {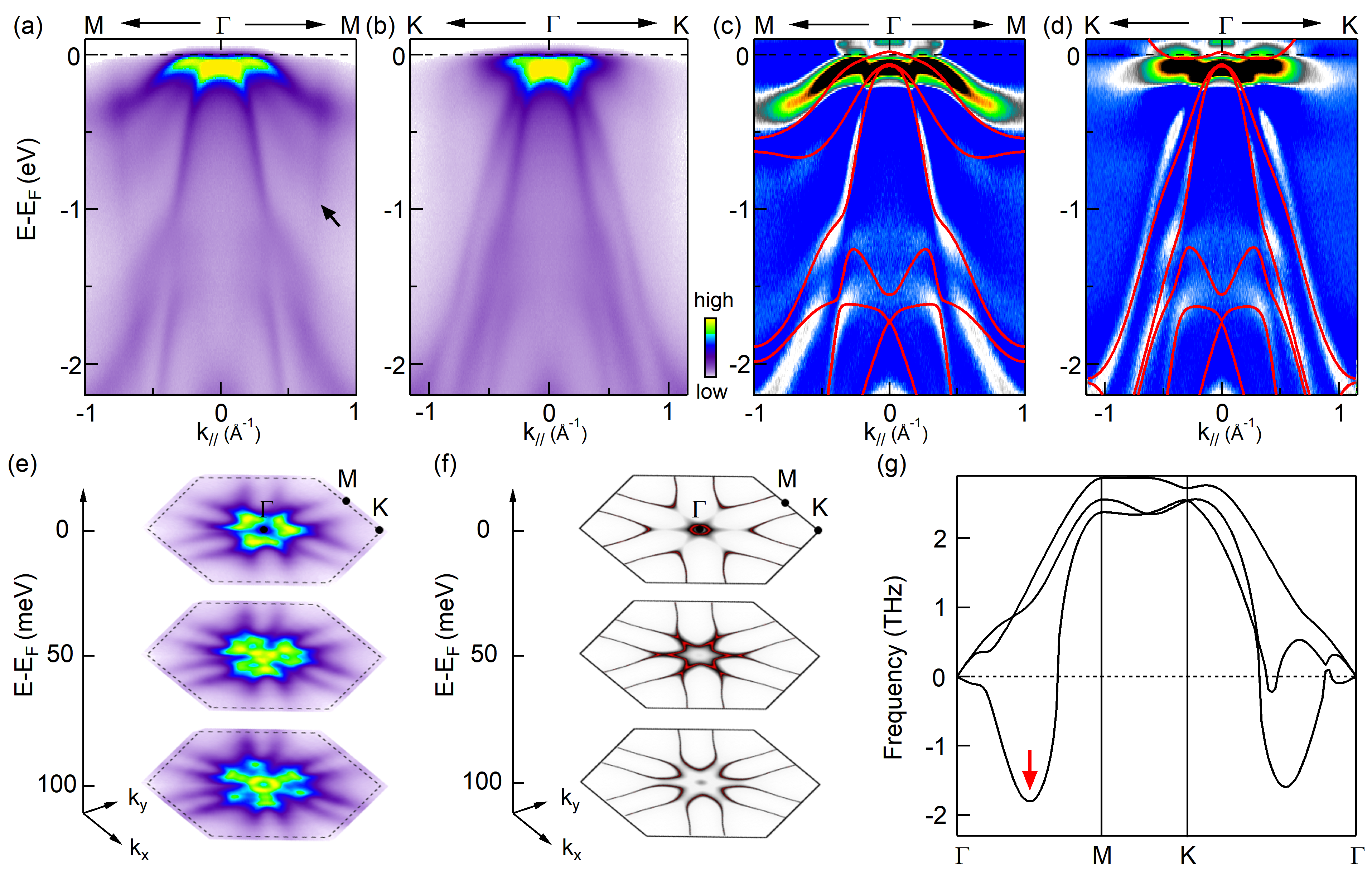}

\caption{ (a) and (b) Band dispersions along the high symmetry directions $\Gamma$-M (a) and $\Gamma$-K (b) with Fermi energy indicated by black dashed line. (c) and (d) Corresponding second derivative of dispersion images for data in (a) and (b). Band structures from VASP are shown in red lines. (e) ARPES intensity maps at 0, -50 meV and -100 meV.  The data were symmetrized by three-fold. The first Brillouin zone (BZ) is indicated by hexagonal dashed lines with high symmetry points labeled. (f) Calculated Fermi surface map by DFT method. (g) Calculated phonon spectrum along the high symmetry directions. The red arrow indicates the softened phonon mode.}
\label{Figure:2}
\end{figure*}

ARPES measurements have been performed to further reveal the electronic structure and the CDW gap. Figures 2(a) and 2(b) show the dispersions measured along two high symmetry directions $\Gamma$-M and $\Gamma$-K, respectively. Near the $\Gamma$ point, there is a rather flat band near the Fermi energy ($\textit{E}_\textup{F}$) from the V 3$\textit{d}$ orbital and a hole pocket from Te 4$\textit{p}$ orbital that extends to -2 eV (See Fig.~S2 in the Supplemental Material \cite{Supplemental}). These two pockets touch at -0.03 eV. Near the M point, there is an electron pocket with the bottom of the band at -0.4 eV. In addition, there are folded bands (pointed by black arrow) which are translated by the high energy Te 4$\textit{p}$ hole pocket at the $\Gamma$ point by a wave vector $\textbf{q}_\textup{1}$ = 1/2 a$^*$ (a$^*$ is the reciprocal lattice vector), suggesting that in addition to the 4$\times$4 reconstruction, there is 2$\times$2 superstructure. Similar 2$\times$2 order has also been observed in other CDW systems, e.g., TiSe$_2$ \cite{ChiangNatcom15}.
Figures 2(c) and 2(d) show the corresponding second derivative of dispersion images along $\Gamma$-M and $\Gamma$-K and the calculated band dispersion (red curve) by VASP for non-magnetic monolayer VTe$_2$. There is an overall agreement between the calculated and experimental band dispersions, suggesting that the 1ML VTe$_2$ film is nonmagnetic. The band structure of monolayer VTe$_2$ is similar to monolayer VSe$_2$ and in sharp contrast with bulk VSe$_2$ and VS$_2$ . For bulk VSe$_2$, the Fermi surface contour consists of six elliptical pockets at the M points and gaps only open at the flat portions \cite{VSe2bulk}. For VS$_2$, the Fermi surface nesting is absent in CDW state, where electron-phonon coupling becomes the dominant driving force \cite{VS2bulk}.

Figures 2(e)-2(f) show a comparison of the ARPES intensity maps at 0, -50 and -100 meV relative to $\textit{E}_\textup{F}$ and the calculated intensity maps. The Fermi surface map shows a circular pocket at the $\Gamma$ point surrounded by six triangular pockets centered at the K points. Below the Fermi energy, the pockets around the K points start to merge with the expanded pocket at the $\Gamma$ point, and they evolve into elliptical hole pockets around the M points. The calculated intensity maps [Fig.~2(f)] are in good agreement with the experimental intensity maps [Fig.~2(e)]. In addition, at $\textit{E}_\textup{F}$, the intensity of the triangular pockets decreases quickly when moving away from the $\Gamma$ point, suggesting possible CDW gap opening near the M point. The calculated phonon spectrum [Fig.~2(g)] clearly indicates that VTe$_2$ is indeed dynamically unstable. The most unstable phonon mode is around the mid-point in the $\Gamma$-M path (indicated by the red arrow), which is consistent with 4$\times$4 CDW order observed in the monolayer film. This suggests that electron-phonon interaction plays an important role in the 4$\times$4 CDW order formation.

 \begin{figure*}
\includegraphics[width=17.8 cm] {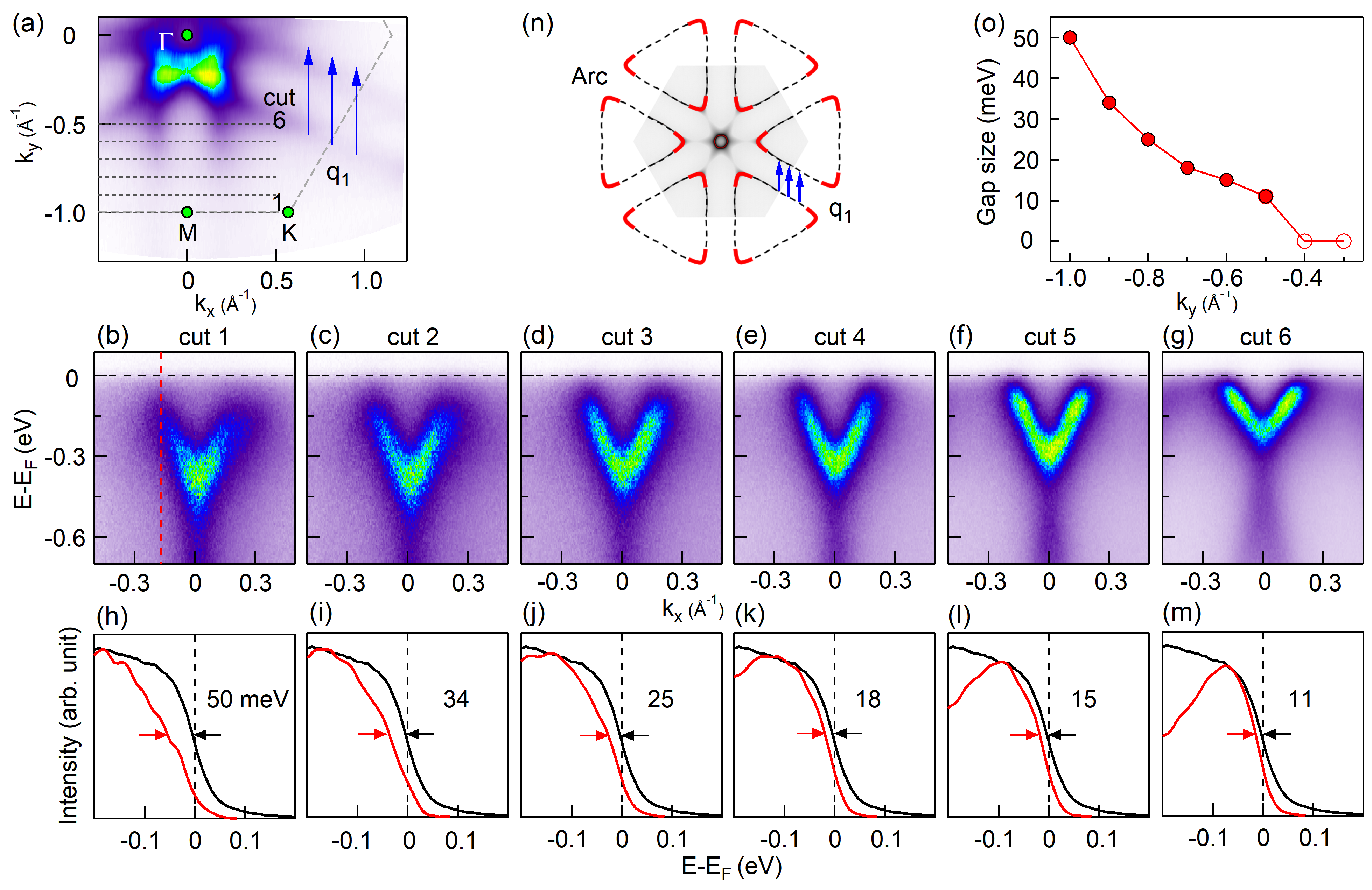}
\caption{(a) Fermi surface map.  Blue arrows indicate $\textbf{q}_\textup{1}$ = 1/4 a$^*$.  (b)-(g) Dispersions along directions parallel to M-K as marked by cuts 1-6 in (a).  (h)-(m) Corresponding EDCs at the Fermi momentum (red lines) to extract the gap size. The black curve is the Fermi edge of polycrystalline gold at 80 K.  (n) Schematic view of pockets around the K points at $\textit{E}_\textup{F}$ and Fermi surface nesting. The black dashed lines and red solid lines represent the gapped region and the arc-like Fermi surface. (o) Extracted gap size.}
\label{Figure:3}
\end{figure*}

To extract the CDW gap, we show a systematic analysis of the electronic structure near $\textit{E}_\textup{F}$ in Fig.~3. Figure~3(a) shows the Fermi surface map. The 4$\times$4 wave vector connects well the straight parts of the triangular pockets centered at the K points, indicating Fermi surface nesting. Figures 3(b)-3(g) show the dispersions measured at a few cuts parallel to the M-K direction indicated by black dashed lines Fig.~3(a). Near the M point (cut 1), there is a strong suppression of intensity near $\textit{E}_\textup{F}$, indicating that there is a gap opening. Moving toward the $\Gamma$ point from cut 1 to cut 6, the suppression of intensity moves toward $\textit{E}_\textup{F}$, indicating decreasing of the gap size.
To quantify the gap size, we show in Figs.~3(h)-3(m) a comparison of the energy distribution curves (EDCs) at the Fermi momentum (red curve) and the Fermi energy reference (black). The CDW gap is extracted from the shift of the EDC leading edge compared to that in the Fermi energy reference and plotted in Fig.~3(o). Here the largest gap size is 50 meV along the M-K direction, and it decreases to 11 meV for cut 6 and eventually vanishes near the $\Gamma$ point (see Fig.~S3 in the Supplemental Material \cite{Supplemental}), forming arc-like Fermi surface. The anisotropic CDW gap is attributed to the imperfect nesting.  As schematically shown in Fig.~3(n), the wave vector $\textbf{q}_\textup{1}$ connects the two parallel parts of the triangular pockets and the gap is the maximum here. Moving toward the $\Gamma$ point, the nesting becomes worse due to the rounded shape near the apex of the triangular pocket, resulting in a reduced gap size. We note that anisotropic CDW gap induced by imperfect nesting is also observed in SmTe$_3$ \cite{GweonPRL98}, CeTe$_3$ \cite{ZXPRL2004} and NbSe$_2$ \cite{NbSePRL}. The energy and momentum resolved capability of ARPES measurements allows us to resolve arc-like Fermi surface and the anisotropic gap due to imperfect nesting of the Fermi surface. Such anisotropic gap is different from that reported in VSe$_2$ where the entire Fermi surface (including the triangular pockets centered at the K points and the circular pockets at the $\Gamma$ point) is gapped \cite{KingNanoLetter18}, suggesting that their CDW mechanisms are different, despite the apparently similar CDW order of 4$\times$4.

\begin{figure*}
\includegraphics[width=17.8 cm] {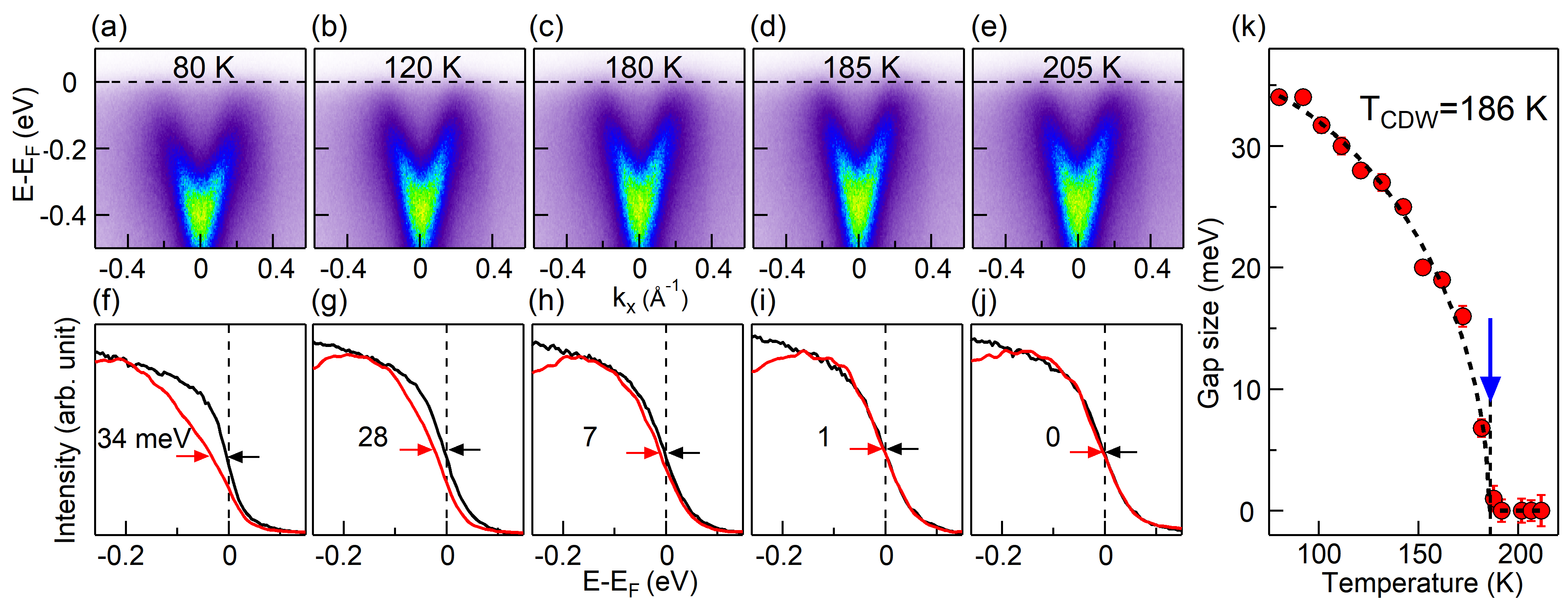}
\caption{(a)-(e) Dispersions measured along the M-K direction at a few selected temperatures. A gap is clearly observed at the top of V 3$\textit{d}$ band in (a). (f)-(j) Corresponding EDCs (red curves) at the gap location for data in (a)-(e) to show the evolution of the gap size. Here the gap value is slightly reduced due to sample degradation after leaving in the ARPES chamber. The black curve is the Fermi energy reference.  (k) Extracted gap size as a function of temperature. Red symbols are the extracted gap size and the black dashed curve is the fitting result using the mean filed theory. From the fitting results, $\textit{T}_\textup{CDW}$ = 186 K.}
\label{Figure:4}
\end{figure*}

Figure 4 shows the evolution of the gap size with temperature. With increasing temperature, the top of the valance band moves gradually to $\textit{E}_\textup{F}$ at higher temperature [Figs. 4(a)-(e)], indicating the decrease of the gap size with temperature. The corresponding EDCs are shown as red curves in Figs.~4(f)-4(j), and the extracted gap size as a function of temperature is shown in Fig.~4(k). The gap size is fitted with mean-field theory \cite{ChiangNatcom15, ZhangKNCuTe}
\begin{equation}\label{1}
  \Delta \propto tanh(A\sqrt{\frac{T_\textup{CDW}}{T}-1})
\end{equation}
 where A is a fitting constant. By fitting the gap size, the CDW transition temperature is extracted to be $\textit{T}_\textup{CDW}$ = 186 $\pm$ 5 K, consistent with the LEED measurements. Taking the maximum gap value of 50 meV and $\textit{T}_\textup{CDW}$ = 186 K, the value of coupling ratio is 2$\Delta/\textit{K}_\textup{B}\textit{T}_\textup{CDW}$ = 6.23. We note that the value of coupling ratio varies in different TMDCs systems, e.g., 6.4 and 17.85 for TaSe$_2$ monolayer \cite{TaSe2CDW,TaSeNanoLett2018}. The extracted ratio for VTe$_2$ film is larger than 3.52 expected for weak-coupled systems, and smaller than the recently reported values of 10 and 38 on monolayer VSe$_2$ \cite{KingNanoLetter18,SatoNanoResearch19}.

We would like to further discuss the driving mechanism of CDW order in monolayer 1T-VTe$_2$ film, and its comparison to CDWs reported in bulk VTe$_2$ as well as VSe$_2$ films. The observation of 4$\times$4 CDW in 1T-VTe$_2$ monolayer film is different from 1T$^\prime$-VTe$_2$ bulk crystal which holds a double zigzag chain-like 3$\times1\times$3 CDW pattern \cite{bronsema1984crystal}. Few-layer 1T-VTe$_2$ films on mica have suggested two possible CDW transitions  \cite{Ma19}, yet the CDW wave vector and gap size still await to be measured. Compared to previous VSe$_2$ film with reported 4$\times$4 CDW order \cite{KingNanoLetter18}, the observation of anisotropic gap opening in our work instead of fully gapped Fermi surface suggests that Fermi surface nesting plays a more important role in the CDW formation of our 1T-VTe$_2$ film. Our ARPES measurements together with phonon spectrum show that both Fermi surface nesting and electron-phonon interaction are the driving forces of CDW formation in monolayer VTe$_2$ films. Similar CDW mechanisms have also been discussed in monolayer 1T-VSe$_2$ film yet with a very different CDW vector $\sqrt{7}\times\sqrt{3}$ \cite{PRL18}, suggesting that there are multiple phonon instabilities in the vanadium dichalcogenide films.
The difference can be attributed to substrate, film thickness or stoichiometry. Further investigation is important to obtain a complete understanding of the evolution of rich CDW physics in vanadium dichalcogenide with thickness and substrates.

In summary, we report the 4$\times$4 CDW order in monolayer 1T-VTe$_2$ film with $\textit{T}_\textup{CDW}$ = 186 K. The observation of anisotropic CDW gap and phonon instability in the calculated phonon spectrum suggest that both Fermi surface nesting and phonon instability contribute significantly to the CDW formation.

This work is supported by the National Natural Science Foundation of China (Grants No.~11725418
and No.~11674188), Ministry of Science and Technology of China (Grants No.~2016YFA0301004, No.~2016YFA0301001, and No.~2015CB921001), Science Challenge Project (No.~TZ2016004) and Beijing Advanced Innovation Center for Future Chip (ICFC).

\end{document}